\documentclass[aps,
reprint,
superscriptaddress,
graphicx,
longbibliography]{revtex4-1}
\usepackage{blindtext}
\usepackage{braket}
\usepackage{graphicx} 
\usepackage[T1]{fontenc}
\usepackage{amsmath,amsthm,amsfonts}
\usepackage{bm}
\usepackage{varioref}
\usepackage{dsfont}
\usepackage{mathbbol}
\usepackage[colorlinks=true,linkcolor=black,citecolor=black]{hyperref}
\usepackage{cleveref}
\title{Sections and Chapters}

\newcommand{\e}{\mathrm{e}}

\newcommand{\Tr}{\mathrm{Tr}}
\renewcommand{\Re}{\mathrm{Re}}
\renewcommand{\Im}{\mathrm{Im}}
\renewcommand{\i}{\mathrm{i}}

\begin{document}
\title{Multiparameter quantum metrology at Heisenberg
scaling for an arbitrary two-channel linear interferometer with squeezed light}
\author{Atmadev Rai}
\affiliation{School of Mathematics and Physics, University of Portsmouth, Portsmouth PO1 3QL, UK}
\author{Danilo Triggiani}
\affiliation{Dipartimento di Fisica, Politecnico di Bari, I-70126 Bari, Italy}
\affiliation{INFN, Sezione di Bari, I-70126 Bari, Italy}
\author{Paolo Facchi}
\affiliation{INFN, Sezione di Bari, I-70126 Bari, Italy}
\affiliation{Dipartimento di Fisica, Universit\`{a} di Bari, I-70126 Bari, Italy}
\author{Vincenzo Tamma}
\email[]{vincenzo.tamma@port.ac.uk}
\affiliation{School of Mathematics and Physics, University of Portsmouth, Portsmouth PO1 3QL, UK}
\affiliation{Institute of Cosmology and Gravitation, University of Portsmouth, Portsmouth PO1 3FX, UK}
\begin{abstract}
We present a framework for simultaneously estimating all four real parameters of a general two-channel unitary $U(2)$ with Heisenberg-scaling precision.
We derive analytical expressions for the quantum Fisher information matrix and show that all parameters attain the $1/N$ scaling in the precision by using experimentally feasible Gaussian probes such as two-mode squeezed states or two single-mode squeezed states. Our results extend multiparameter metrology to its most general two-mode setting and establish concrete design principles for experimental implementations of Heisenberg-scaling, multi-parameter optical interferometry with experimentally feasible resources.  It not only sheds light on the fundamental interface between quantum interference of squeezed light and quantum metrological advantage in multiparameter estimation, but it also provides an important stepstone towards the development of a wide range of quantum technologies based on distributed quantum metrology in arbitrary optical networks.
\end{abstract}

\maketitle
\section{Introduction}
Quantum metrology harnesses nonclassical resources to improve measurement precision beyond classical limits. Conventional interferometers are described by $SU(2)$ transformations, and with appropriately entangled or squeezed probe states with $N$ average number of photons, these setups can in principle attain phase sensitivities with the Heisenberg limit $\Delta\Phi\sim 1/N$~\cite{PhysRevD.23.1693,PhysRevLett.71.1355,PhysRevD.30.2548,PhysRevLett.96.010401,dowling2015quantum,zhou2018achieving,qian2019heisenberg,doi:10.1126/science.1104149,PhysRevLett.102.100401,PhysRevA.92.022106}.

In single-parameter estimation, one often achieves the so-called Heisenberg scaling with optimal entangled probes. Recent work has extended these ideas to the simultaneous estimation of multiple parameters, showing that joint strategies can outperform independent one-parameter strategies. Extensive theoretical~\cite{PhysRevLett.111.070403, imai2007geometry, PhysRevLett.116.030801, szczykulska2016multi, PhysRevA.94.052108} and experimental~\cite{polino2019experimental,hou2016achieving, PhysRevLett.124.140501, roccia2017entangling} studies have addressed increasingly sophisticated multiparameter problems, including the simultaneous estimation of phase and diffusion noise~\cite{vidrighin2014joint, PhysRevA.92.032114}, phase
and imperfection in the probe state~\cite{Roccia:18}, two-phase estimation with Heisenberg scaling~\cite{PhysRevA.111.062408}.

In quantum metrology, estimating multiple parameters simultaneously involves multiple uncertainty relations; however, the interplay among multiple uncertainties remains largely unexplored \cite{ALBARELLI2020126311,PhysRevA.98.012114,PhysRevX.12.011039,chang2025multiparameter}. The key challenge stems from two main incompatibilities: First, measurement incompatibility, where the quantum Cramér-Rao bound may not be saturated for all parameters simultaneously~\cite{KMatsumoto_2002,PhysRevA.89.023845,vidrighin2014joint}. Second, probe incompatibility, where a single probe state may not offer maximal sensitivity for all parameters~\cite{PhysRevX.12.011039}. These trade-offs mean that achieving the absolute Heisenberg limit for each parameter simultaneously may not be possible. Instead, a more practically relevant goal is to achieve Heisenberg-scaling precision $\Delta\Phi\propto 1/N$ with experimentally accessible states and measurements. The central motivation of this study is to address this gap by deriving simultaneous precision bounds for all parameters of interest in an arbitrary two-channel linear interferometer. In doing so, we not only map out the fundamental scaling in the precision of multiparameter estimation but also shed new light on the structure and interplay of the Gaussian input probe.

In this article, we analyze the simultaneous estimation of all four real unknown parameters defining an arbitrary two-channel linear-optical transformation. A generic $U(2)$ transformation on two optical modes can be parameterized by four real angles, which encapsulates all possible linear optical transformations, including beam splitters, phase shifters, and mode-mixing operations, on a pair of bosonic modes. A widely used tool that characterizes the ultimate precision in quantum metrology is the quantum Fisher information matrix (QFIM). We derive the full QFIM for all four parameters of the unitary injected by the Gaussian input probe. We first consider a two-mode squeezed state (TMSS) as an input. We show that the quantum Cramér-Rao bound (QCRB) achieves the Heisenberg scaling for all four parameters simultaneously for a given fixed average number of input photons $N$. The second input state we consider is two single-mode squeezed states (SMSSs); we evaluate the QCRB for all four parameters for the optimal phase-matching condition of the input probe. We show that, by optimally tuning the displacement and squeezing of either the TMSS or two SMSS probes, one can achieve the Heisenberg scaling in the simultaneous estimation of all four parameters. This study thus extends interferometric quantum metrology into a full multi-parameter regime by tackling the simultaneous estimation of the set of physical parameters with a scalable Gaussian probe and experimentally feasible measurements.

\section{PRELIMINARIES}
\subsection{Arbitrary two-channel linear interferometer}
\label{SectionII}
\begin{figure}
    \centering
    \includegraphics[width=0.75\linewidth]{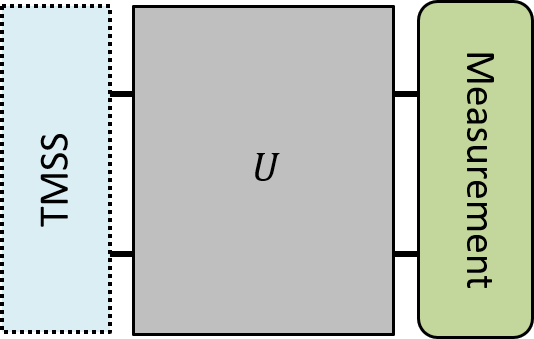}
  \caption{Schematic of the estimation setup. A TMSS, in Eq.~\eqref{eq:INPUTstate} is injected into an unknown two-channel unitary $U$.}
    \label{fig1}
\end{figure}
In \figurename~\ref{fig1} we consider an arbitrary two-channel linear, passive transformation that includes various optical and atomic passive operations, such as beam splitters, Mach-Zehnder, and Ramsey interferometers~\cite{PhysRevA.50.67}. The input and output mode annihilation operators are related by 
\begin{equation}
    \begin{pmatrix}
        \hat{a}_{1,\mathrm{out}} \\ \hat{a}_{2,\mathrm{out}}
    \end{pmatrix}=U \begin{pmatrix}
        \hat{a}_{1,\mathrm{in}} \\ \hat{a}_{2,\mathrm{in}}
    \end{pmatrix}.
\end{equation}
The transformation matrix $U$ can be written as
\begin{equation}
    U=\e^{i\phi_0}\begin{pmatrix}
       \e^{i\phi_\tau}\cos{\omega}&\e^{i\phi_\rho}\sin{\omega} \\-\e^{-i\phi_\rho}\sin{\omega}&\e^{-i\phi_\tau}\cos{\omega}
    \end{pmatrix},
    \label{eq:groupU(2)}
\end{equation}
which represents the most general unitary transformation and belongs to $U(2)$ with determinant $\e^{2i\phi_0}$ and preserves the bosonic commutation relations.
The parameter $\omega$ is physically related to the transmittance $\tau=\cos^2{\omega}$ and reflectance $\rho=\sin^2{\omega}$ of the transformation with the corresponding phases $\phi_t$ and $\phi_r$, respectively. The general transformation in Eq.~\eqref{eq:groupU(2)} can also be expressed as the product of four matrices as~\cite{PhysRevA.40.1371}
\begin{equation}
\begin{split}
 U= \begin{pmatrix}
        \e^{i\phi_0}&0\\0&\e^{i\phi_0}
\end{pmatrix}\times\begin{pmatrix}\e^{i\psi/2}&0\\0&\e^{-i\psi/2} 
    \end{pmatrix}\\\times \begin{pmatrix}
        \cos{\omega}&\sin{\omega}\\-\sin{\omega}&\cos{\omega}
    \end{pmatrix}\times \begin{pmatrix}
        \e^{i\phi/2}&0\\0&\e^{-i\phi/2}
    \end{pmatrix} , 
\end{split}
\label{eq:unitarymatrix}
\end{equation}
where $\phi_\tau=(\phi+\psi)/2$ and $\phi_\rho=(\phi-\psi)/2$. The parameters take values in the ranges $\phi_0,\phi,\psi\in[0,2\pi)$, $\omega\in[0,\pi/2]$.

To discuss the quantum properties, it is convenient to write the two-mode unitary $\hat{U}$, corresponding to Eq.~\eqref{eq:unitarymatrix} thru $\hat{a}_{k,\mathrm{out}}=\hat{U}\hat{a}_{k,\mathrm{in}}\hat{U}^\dagger$, with $k=1,2$, in terms of the Schwinger representation of angular momentum operators as~\cite{schwinger1965quantum}
\begin{equation} \hat{U}=\e^{i\phi_0\hat{N}}\e^{i\psi\hat{J}_z}\e^{i\omega\hat{J}_y}\e^{i\phi\hat{J}_z},
\label{eq:U}
\end{equation}
where $\hat{N}=\hat{a}_1^\dagger\hat{a}_1+\hat{a}_2^\dagger\hat{a}$ is the number of particle operator, and 
\begin{equation*}
\begin{split}
\hat{J}_x=\frac{\hat{a}_1^\dagger\hat{a}_2+\hat{a}_2^\dagger\hat{a}_1}{2},\\\hat{J}_y=\frac{\hat{a}_1^\dagger\hat{a}_2-\hat{a}_2^\dagger\hat{a}_1}{2\i},\\
\hat{J}_z=\frac{\hat{a}_1^\dagger\hat{a}_1-\hat{a}_2^\dagger\hat{a}_2}{2}
\end{split} 
\end{equation*}
are the usual angular momentum operators. The generators associated with the parameters are defined as $\hat{G}_k=-\i \hat{U^\dagger}\partial_k\hat{U}$ and read
\begin{equation}
\begin{split}
    \hat{G}_{\phi_0}&=\hat{N},\\
    \hat{G}_{\phi}&=\hat{J}_z,\\
    \hat{G}_{\psi}&=\hat{J}_z\cos{\omega}+(\hat{J}_x\cos{\phi}+\hat{J}_{y}\sin{\phi})\sin{\omega},\\
    \hat{G}_{\omega}&=\hat{J}_y\cos{\phi}-\hat{J}_x\sin{\phi}.
    \end{split}
    \label{eq:Generators}
\end{equation}
We note that, for the unitary $\hat{U}$ in Eq.~\eqref{eq:U}, the phase shift $\phi_0$ associated with the operator $\e^{i\phi_0\hat{N}}$ becomes irrelevant if both the input state and the POVM do not contain coherent superpositions of quantum states with different numbers of photons making the output probabilities insensitive to the parameter $\phi_0$. However, if the number coherences are present in both the probe state and POVM (e.g., in a homodyne measurement), the parameter $\phi_0$ can be estimated~\cite{PhysRevA.91.032103}. 

\subsection{Gaussian states}
In this work, we restrict our discussion to Gaussian states. 
The Wigner distribution function of a two-mode Gaussian state
\begin{equation}
W(\bm{z}) = \frac{1}{(2\pi)^2 \sqrt{\det[\Gamma]}} \exp \left( -\frac{1}{2} (\bm{z} - \bm{d})^T \Gamma^{-1} (\bm{z} - \bm{d}) \right),
\end{equation}
with $\bm{z} \in \mathbb{R}^{4}$ being the phase-space coordinate vector, is fully characterized by $\bm{d}$ and $\Gamma$, the first and second moments of the Gaussian state, respectively. The real symplectic matrix of a linear, passive optical two-mode transformation \(U\) on  phase-space is given by
\begin{equation}
R = \left(\begin{array}{cc}
 \Re[U] & -\Im[U] \\
\Im[U] & \Re[U]
\end{array}\right).
\label{EqnA3}
\end{equation}
Under the transformation $U$, Gaussian states remain Gaussian, with the displacement vector and covariance matrix transforming as $\bm{d}_U=R\,\bm{d}$ and $\Gamma_U =R\,\Gamma\, R^{T}$, respectively.

\section{Multiparameter quantum Cramer-Rao bound}
\label{SectionIII}
Our goal is to estimate all physical parameters in an arbitrary two-channel interferometer with ultimate precision, using experimentally feasible and robust Gaussian states.
The precision of any locally unbiased estimator $\tilde{\bm{\Phi}}$ of the parameter vector $\bm{\Phi}$ is fundamentally bounded by the QFIM $\mathcal{H}$. In particular, the QCRB 
\begin{equation}
\mathrm{Cov}(\tilde{\bm{\Phi}})\geq \frac{1}{M}\mathcal{H}^{-1},
\label{eq:QCRB}
\end{equation}
where $\mathrm{Cov}(\tilde{\bm{\Phi}})$ is the error covariance matrix associated with  the estimator $\tilde{\bm{\Phi}}$ and $M$ is the number of measurement iterations, sets the lower bound on the achievable sensitivities. For simplicity, throughout this paper, we set $ M=1$. The matrix elements of $\mathcal{H}$ can be written directly in terms of the generators~\eqref{eq:Generators} as
\begin{equation}
    \mathcal{H}_{ij}=4(\braket{\hat{G}_i\hat{G}_j}-\braket{\hat{G}_i}\braket{\hat{G}_j})
\end{equation}
where we denote $\braket{\hat{O}}=\bra{\psi_{\mathrm{in}}}\hat{O}\ket{\psi_{\mathrm{in}}}$, with $\ket{\psi_{\mathrm{in}}}$ being the initial state.

For the Gaussian state characterized by the displacement vector $\bm{d}_U$ and covariance matrix $\Gamma_U$, we can write the QFIM as
\begin{equation}
    \mathcal{H}=\mathcal{H}^{\bm{d}}+\mathcal{H}^\Gamma,
    \label{eq:QFIMmain}
\end{equation}
with matrix elements
\begin{equation}
\begin{split}
    \mathcal{H}^{\bm{d}}_{mn}&=\partial_m \bm{d}^{T}_U \Gamma^{-1}_U \partial_n \bm{d}_U,\\
    \mathcal{H}^\Gamma_{mn}&=\frac{1}{2} \mathrm{Tr}\left[\Gamma_U^{-1} (\partial_m\Gamma_U) \Gamma^{-1}_U (\partial_n \Gamma_U)\right],
    \end{split}
\end{equation}
where $\mathrm{Tr}[.]$ denotes the trace operation. The QFIM in Eq.~\eqref{eq:QFIMmain} decomposes into the sum of two terms $\mathcal{H}^\Gamma$ and $\mathcal{H}^{\bm{d}}$ arising from parameter variations of the covariance matrix $\Gamma_U$ and the displacement vector $\bm{d}_U$, respectively.

We will show that the TMSS allows Heisenberg scaling precision in the estimation of all unknown parameters in the two-mode unitary.

\subsection{Two-mode squeezed state}
We evaluate the QFIM for all four parameters for the input TMSS, defined by
\begin{equation}
\ket{\psi_{\mathrm{in}}}=\hat{D}_1(\alpha_1)\hat{D}_2(\alpha_2)\hat{S}_{12}(\xi)\ket{0,0},
\label{eq:INPUTstate}
\end{equation}
where the $\hat{D}_i(\alpha_i)=\exp(\alpha_i \hat{a}_i^\dagger - \alpha_i^* \hat{a}_i)$ is the displacement operator at the $i$-th input port with $\alpha_i=|\alpha_i|\e^{\i\beta_i}$ and $\hat{S}_{12}(\xi)=\exp{(\xi^*\hat{a}_1\hat{a}_2-\xi\hat{a}^\dagger_1\hat{a}^\dagger_2)}$ is the operator of the two-mode squeezing with $\xi=r\e^{i\theta}$. The first and second moments of the state can be expressed as
\begin{equation}
\bm{d}
=
\sqrt{2}\Big(\Re(\alpha_1),\;\Im(\alpha_1),\;\Re(\alpha_2),\;\Im(\alpha_2)\Big),
\end{equation}
\begin{equation}
   \Gamma_{\rm TMSS}
= \frac{1}{2}\begin{pmatrix}
   \cosh{{2r}}\, \mathbb{1}_2&-\sinh{2r}\, S(\theta)\\
  - \sinh{2r}\, S(\theta)&\cosh{{2r}}\, \mathbb{1}_2
\end{pmatrix},
\label{eq:TMSSCOV}
\end{equation}
where $\mathbb{1}_2$ is $2\times2$ identity matrix and
\begin{equation}
    S(\theta)=\begin{pmatrix}
        \cos{\theta}&\sin{\theta}\\\sin{\theta}&-\cos{\theta}
    \end{pmatrix}.
\end{equation}
\begin{figure}
    \centering
    \includegraphics[width=\linewidth]{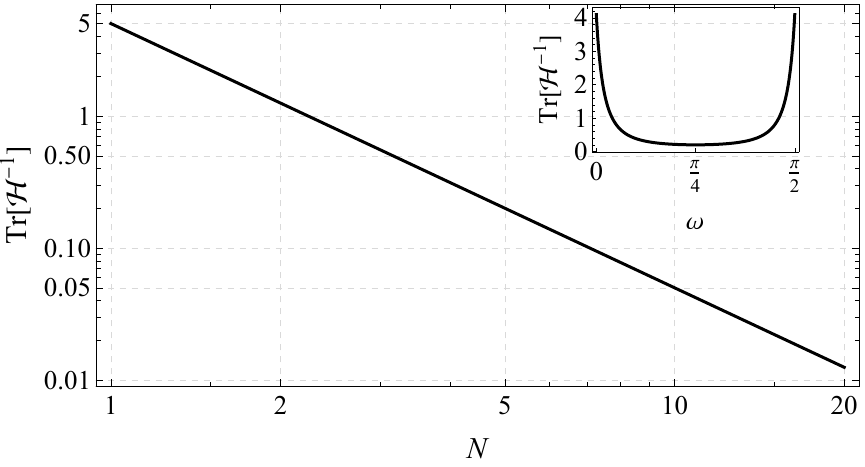}
    \caption{Log--log plot of the scalar precision bound \(\Tr[\mathcal{H}^{-1}]\) (sum of the QCRBs), as a function of the total average photon number \(N\). The scalar bound \(\Tr[\mathcal{H}^{-1}]\) scales as \(1/N^2\), demonstrating Heisenberg-scaling precision in all parameters. The inset plot shows the dependence of \(\mathrm{Tr}[\mathcal{H}^{-1}]\) on the parameter \(\omega\) at fixed $N=5$, illustrating the parameter dependency of the total precision bound. Here, we choose the optimal value $\omega=\pi/4$ for the main plot.  
}
    \label{fig:figure2}
\end{figure}
A direct calculation of $\mathcal{H}^\Gamma$ in Eq.~\eqref{eq:QFIMmain}, describing the contribution to the sensitivity due to variations of the covariance matrix $\Gamma$,  gives
\begin{equation}
   \mathcal{H}_{\rm TMSS}^\Gamma =\begin{pmatrix}
    8\sinh^2{2r}&0&0&0\\0&0&0&0\\0&0&2\sinh^2{2r}\sin^2{2\omega}&0\\0&0&0&8\sinh^2{2r}
    \end{pmatrix}.
    \label{eq:QFIM_Gamma_TMSS}
\end{equation}
It is evident from $\mathcal{H}^\Gamma_{\rm TMSS}$ in Eq.~\eqref{eq:QFIM_Gamma_TMSS} that the two-mode squeezed vacuum alone is not enough to estimate all four parameters, as the QFI for the parameter $\phi$ is zero. This outcome is physically intuitive as the parameter $\phi$ is associated with the generator $\hat{G}_\phi=\hat{J}_z$ in Eq.~\eqref{eq:Generators}, which corresponds to the difference in photon number between the two modes and leads to vanishing variance $\mathrm{Var}(\hat{J}_z)=0$ for a two-mode squeezed vacuum state.

We then compute the contribution to the QFIM arising from variations of the displacement vector, denoted by $\mathcal{H}^{\bm{d}}$ in Eq.~\eqref{eq:QFIMmain}. A detailed derivation of $\mathcal{H}^{\bm{d}}$ is given in Appendix~\ref{Appendix A}. Introducing displacement at the input yields a nonzero Fisher information for the previously inaccessible parameter $\phi$ and thus enables estimation of all four parameters at Heisenberg scaling. Moreover, $\mathcal{H}^{\bm{d}}$ contributes at most two eigenvalues scaling as $O(N^2)$,
with $N$ being the average number of input photons, therefore at most two independent parameter combinations can achieve Heisenberg scaling. By optimally tuning the input squeezing and displacement phases to $\theta-\beta_1-\beta_2=0$ and by choosing equal displacement amplitudes $|\alpha_{1(2)}|=|\alpha|$, the Fisher information for $\phi$ is maximized, taking the simple form $\mathcal{H}^{\bm{d}}_{22}=2|\alpha|^2\e^{2r}$, and the scalar bound $\Tr[\mathcal{H}_{\rm TMSS}^{-1}]$ (the sum of the QCRBs) is minimized (see \figurename~\ref{fig:displacement_weight} in Appendix~\ref{Appendix A}).

Writing the mean photon numbers contributed by squeezing and displacement as \(N_s=2\sinh^2 r\) and \(N_c=N_{c1}+N_{c2}\) with \(N_{c_{1(2)}}=|\alpha_{1(2)}|^2\), so that \(N=N_s+N_c\) be the total average number of photons in the TMSS, the total QFIM becomes (see Appendix~\ref{Appendix A})
\begin{equation}
\mathcal{H}_{\rm TMSS} =
\begin{pmatrix}
8N_s^2 & 0 & 0 & 0 \\
0 & N_cN_s & N_cN_s\cos{2\omega} & 0 \\
0 & N_cN_s\cos{2\omega} &
\begin{array}{@{}c@{}}
\big(2N_s^2\sin^2{2\omega}\\
+\,N_cN_s\cos^2{2\omega}\big)
\end{array}
& 0 \\
0 & 0 & 0 & 8N_s^2
\end{pmatrix},
\label{eq:QFTMSS}
\end{equation}
where we retain only leading terms of order $O(N^2)$, with $N_{c,s}=O(N)$, since lower-order terms do not contribute to the Heisenberg scaling for large $N$. From the QFIM in Eq.~\eqref{eq:QFTMSS} we arrive at the QCRB matrix inequality~\eqref{eq:QCRB}, which implies that
\begin{equation}
\begin{split}
(\Delta\phi_0)^2&\geq(\mathcal{H}^{-1})_{11}=\frac{1}{8 N_s^2},\\
(\Delta\phi)^2&\geq(\mathcal{H}^{-1})_{22}=\frac{\frac{2N_s}{N_c}+\cot^2{2\omega}}{2 N_s^2},\\
(\Delta\psi)^2&\geq(\mathcal{H}^{-1})_{33}=\frac{1}{2 N_s^2\sin^2{2\omega}},\\
(\Delta\omega)^2&\geq(\mathcal{H}^{-1})_{44}=\frac{1}{8 N_s^2}.\\
\end{split}
\label{eq:QCRB4parameter}
\end{equation}
These bounds confirm that all four parameters can be estimated simultaneously with the Heisenberg-scaling sensitivity. \figurename~\ref{fig:figure2} presents the scalar bound $\Tr[\mathcal{H}_{\rm TMSS}^{-1}]$ by considering equal photons in squeezing and displacement, i.e., $N_s=N_c=N/2$, as a function of total average number of photons $N$ and the parameter $\omega\in(0,\pi/2)$.

\subsection{Two single-mode squeezed states}
We now consider the squeezed displaced plus squeezed displaced combination into the input channels, given by
\begin{equation}
    \ket{\psi_{in}}=\ket{\xi_1\alpha_1}_1\otimes\ket{\xi_2\alpha_2}_2
    \label{eq:SMSS}
\end{equation}
where $\ket{\xi_i\alpha_i}_i=\hat{D_i}(\alpha_i)\hat{S}_i(\xi_i)\ket{0}_i$. Here, $\hat{S}_i(\xi_i)$ is the single-mode squeezing operator with parameter $\xi_i=r_i \e^{2\i\theta_i}$, and $\hat{D_i}(\alpha_i)$ is the displacement operator acting on mode $i=1,2$. For simplicity, we consider here the case of equal squeezing parameter $r_1=r_2=r$, although the obtained result can be generalized to arbitrary $r_1$ and $r_2$ given in Appendix~\ref{Appendix B}. We show in Appendix~\ref{Appendix B} that the QFIM for SMSSs ($\mathcal{H}_{\rm SMSS}$) is directly analogous to the TMSS case, as the squeezing alone allows us to estimate only three of four parameters with Heisenberg scaling, and displacement at the input remains necessary to estimate all four parameters with Heisenberg scaling. However, in contrast to the TMSS case, the QFIM $\mathcal{H}^\Gamma_{\rm SMSS}$ is not diagonal and leads to a cumbersome expression of the QCRB.

\begin{figure}
    \centering
    \includegraphics[width=0.8\linewidth]{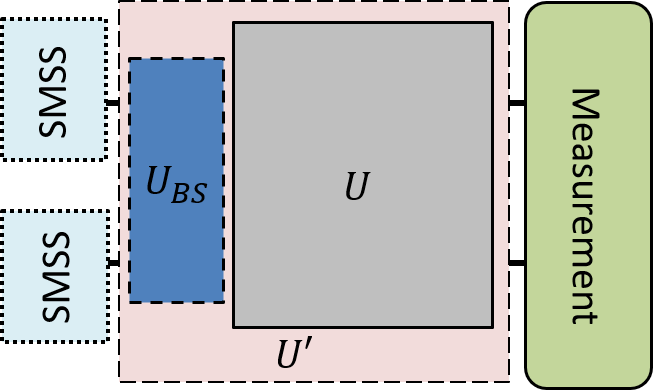}
    \caption{A $50{:}50$ BS transformation $U_{\rm BS}$ is applied to a general unitary operation. While $U_{\rm BS}$ is fixed, the overall unitary remains arbitrary, preserving the generality of the transformation.}
    \label{fig:3}
\end{figure}

Therefore, for analytical convenience and to make the calculations less cumbersome, we choose to rewrite the unitary by absorbing a fixed $50{:}50$ beam-splitter (BS) transformation $U_{\rm BS}$ into the unitary $U$. In \figurename~\ref{fig:3}, the total unitary can be written as
\begin{equation}
    U' = U\,U_{\mathrm{\rm BS}}.
\end{equation}
Equivalently, we can also think of the $U_{\mathrm{BS}}$ as part of the probe preparation, such as the SMSS together with BS defines the effective input state. In particular, two SMSSs with equal squeezing parameter incident on $50{:}50$ BS and conditioning on the relative phase $\theta_1-\theta_2=\pm\pi/2$ yield a TMSS \cite{PhysRevLett.131.193601}
\begin{equation}
    \Gamma_{\rm TMSS}\;=\;R(U_{\rm BS})\;\Gamma_{\rm SMSS}\;R(U_{\rm BS})^T,
\end{equation}
where $R$ being the symplectic rotation defined in Eq.~\eqref{EqnA3}.
We emphasize that this additional BS rotation is not necessary to achieve the Heisenberg-scaling precision in the estimation of all four parameters, as shown in Appendix~\ref{Appendix B}; it is introduced only to reduce the complexity. One can simply find that the total QFIM for the SMSSs after the BS rotation, $\mathcal{H}_{\rm SMSS}[\{U'\}]$, coincides with the QFIM for the TMSS $\mathcal{H}_{\rm TMSS}[\{U\}]$ for the same total mean photon number $N$ at the input. In other words,
\begin{equation}
    \mathcal{H}^\Gamma_{\mathrm{SMSS}}[\{U' \}] = \mathcal{H}^\Gamma_{\mathrm{TMSS}}[\{ U \}],
\end{equation}
and for equal displacement amplitude $|\alpha_1|^2=|\alpha_2|^2=N_c/2$ and phases fixed as $\beta_1=\pi/2$ and $\beta_2=0$, the total QFIM asymptotically reads
\begin{equation}
\mathcal{H'}_{\rm SMSS} = 
\begin{pmatrix}
8N_s^2 & 0 & 0 & 0 \\
0 & N_cN_s & N_cN_s\cos{2\omega} & 0 \\
0 & N_cN_s\cos{2\omega} & \begin{array}{@{}c@{}}
\big(2N_s^2\sin^2{2\omega}\\
+\,N_cN_s\cos^2{2\omega}\big)
\end{array} & 0 \\
0 & 0 & 0 & 8N_s^2 \\
\end{pmatrix}.
\label{eq:QFSMSS}
\end{equation}
From Eq.~\eqref{eq:QFTMSS} we can see that $\mathcal{H'}_{\rm SMSS}=\mathcal{H}_{\rm TMSS}$ where $\mathcal{H'}$ represents the QFIM corresponding to the effective unitary $U'$. Therefore, the QCRB derived from this analysis matches exactly the result in Eq.~\eqref{eq:QCRB4parameter}.

\section{Conclusions}
In this work, we propose a comprehensive framework for multiparameter estimation of an arbitrary two-mode linear transformation $U(2)$ at Heisenberg-scaling precision. We have analyzed two experimentally feasible classes of Gaussian probe states, a TMSS and two SMSSs, for the simultaneous estimation of all four physical unknown parameters of an arbitrary two-channel interferometer, each achieving the Heisenberg scaling in the sensitivity by exploiting the QCRB. We have shown that only squeezing in the two input modes, such as two-mode squeezed vacuum state and two single-mode squeezed vacuum states, is not enough to estimate all four parameters, and displacement in the probe is necessary to achieve the Heisenberg-scaling precision for each parameter simultaneously given by the QCRB in Eq.~\eqref{eq:QCRB4parameter}. Our results based on
the QCRB set a useful benchmark for designing experiments that lead to Heisenberg scaling in the estimation of all parameters in arbitrary two-channel linear interferometers and will pave the way to future works extending our results to a larger number of channels. This work represents therefore an important stepstone towards achieving the ultimate quantum precision in the characterization of arbitrary linear optical interferometers with experimentally feasible resources. This will inform the development of scalable quantum technologies for distributed quantum metrology with important applications, ranging from characterization of quantum gates, quantum imaging, biomedical sensing, nonlinear functions interpolation, and estimation of gravitational wave parameters, as well as inhomogeneous forces and gradients.

\acknowledgments

This work was partially supported by Xairos System Inc. VT also acknowledges partial support from the Air Force Office of Scientific Research under award number FA8655-23-17046.
PF was partially supported by Istituto Nazionale di Fisica Nucleare (INFN) through the project ``QUANTUM'', by the Italian National Group of Mathematical Physics (GNFM-INdAM), and by the Italian funding within the ``Budget MUR - Dipartimenti di Eccellenza 2023--2027'' - Quantum Sensing and Modelling for One-Health (QuaSiModO). D.T. acknowledges the Italian Space Agency (ASI, Agenzia Spaziale Italiana) through the project Subdiffraction Quantum Imaging (SQI) No.2023-13-HH.0.

\newpage
\onecolumngrid
\appendix
\numberwithin{equation}{section}
\renewcommand{\theequation}{A\arabic{equation}}  
\section{Quantum Fisher Information Matrix for Input TMSS} \label{Appendix A}
\numberwithin{equation}{section}
\renewcommand{\theequation}{A\arabic{equation}}       
  \setcounter{equation}{0} 
In this appendix, we obtain the QFIM for all four physical parameters using TMSS. Consider the input state
\begin{equation}
\ket{\psi_{\mathrm{in}}}
= \hat{D}_1(\alpha_1)\,\hat{D}_2(\alpha_2)\,\hat{S}_{12}(\xi)\,\ket{0,0},
\label{App:INPUTstate}
\end{equation}
where $\hat{D}_i(\alpha_i)$ is the single‐mode displacement operator on mode $i$ with $\alpha_i=|\alpha_i|\e^{i\beta_i}$ for $i=1,2$, and $\hat{S}_{12}(\xi)$ is the two‐mode squeezing operator with complex squeezing parameter $\xi = r\,e^{i\theta}$.  In the following, we will adopt the quadrature ordering 
\(\mathbf{z}=(x_1,\,p_1,\,x_2,\,p_2)^{T}\).

We can write the initial displacement vector of the input state~\eqref{App:INPUTstate} as
\begin{equation}
\bm{d}
=
\sqrt{2}\begin{pmatrix}
\,|\alpha_1|\cos{(\beta_1)} \\
\,|\alpha_1|\sin{(\beta_1)} \\
\,|\alpha_2|\cos{(\beta_2)} \\
\,|\alpha_2|\sin{(\beta_2)}
\end{pmatrix},
\label{App:intial_d}
\end{equation}
since each displacement \(\hat{D}_i(\alpha_i)\) shifts \(\langle x_i\rangle\) by \(\sqrt{2}\,\Re(\alpha_i)\) and \(\langle p_i\rangle\) by \(\sqrt{2}\,\Im(\alpha_i)\).

The TMSS is then injected into a two-mode passive, linear interferometer described by a unitary matrix $U$, which reads
\begin{equation}
U 
= 
\begin{pmatrix}
e^{\,i\phi_0} & 0 \\
0 & e^{\,i\phi_0}
\end{pmatrix}
\;\times\;
\begin{pmatrix}
e^{\,i\psi/2} & 0 \\
0 & e^{\,-\,i\psi/2}
\end{pmatrix}
\;\times\;
\begin{pmatrix}
\cos\omega & \sin\omega \\
-\,\sin\omega & \cos\omega
\end{pmatrix}
\;\times\;
\begin{pmatrix}
e^{\,i\phi/2} & 0 \\
0 & e^{\,-\,i\phi/2}
\end{pmatrix}.
\label{eq:U2_decomposition}
\end{equation}
here, $\phi_0, \phi, \psi\in[0,2\pi)$, and $\omega\in[0,\pi/2]$ are the four real parameters we aim to estimate. In the phase-space formalism, $U$ acts as a symplectic rotation 
\begin{equation}
    R = \left(\begin{array}{cc}
 \Re[U] & -\Im[U] \\
\Im[U] & \Re[U]
\end{array}\right),
\label{App:R}
\end{equation}
so that the transformed first and second moments read
\begin{equation}
    \bm{d}_U= R\,\bm{d}=\sqrt{2}
\begin{pmatrix}
& 
|\alpha_1| \cos{\omega} \cos\left( \beta_1 + \frac{\psi + \phi}{2} + \phi_0 \right)
+ |\alpha_2| \sin{\omega} \cos\left( \beta_2 + \frac{\psi - \phi}{2} + \phi_0 \right)
\\
&
|\alpha_1| \cos{\omega} \sin\left( \beta_1 + \frac{\psi + \phi}{2} + \phi_0 \right)
+ |\alpha_2| \sin{\omega} \sin\left( \beta_2 + \frac{\psi - \phi}{2} + \phi_0 \right) \\
&
|\alpha_2| \cos{\omega} \cos\left( \beta_2 - \frac{\psi + \phi}{2} + \phi_0 \right)
- |\alpha_1| \sin{\omega} \cos\left( \beta_1 - \frac{\psi-\phi}{2}  + \phi_0 \right) \\
& 
|\alpha_2| \cos{\omega} \sin\left( \beta_2 - \frac{\psi + \phi}{2} + \phi_0 \right)
- |\alpha_1| \sin{\omega} \sin\left( \beta_1 - \frac{\psi-\phi}{2} + \phi_0 \right)
\
\end{pmatrix},
\label{App:evolved_d}
\end{equation}
\begin{equation}
\begin{aligned}
 \Gamma_U=R\,\Gamma_{\rm TMSS}\,R^T
&=\frac{1}{2} 
\left(\begin{matrix}
\cosh{2r} - \sin{2\omega}\sinh{2r}\cos(\theta + \psi + 2\phi_0) &
- \cos{2\omega}\sinh{2r}\cos(\theta + 2\phi_0)\\
- \cos{2\omega}\sinh{2r}\cos(\theta + 2\phi_0) &
\cosh{2r} + \sin{2\omega}\sinh{2r}\cos(\theta - \psi + 2\phi_0) \\- \sin{2\omega}\sinh{2r}\sin(\theta + \psi + 2\phi_0) &
- \cos{2\omega}\sinh{2r}\sin(\theta + 2\phi_0) \\- \cos{2\omega}\sinh{2r}\sin(\theta + 2\phi_0) &
\sin{2\omega}\sinh{2r}\sin(\theta - \psi + 2\phi_0)
\end{matrix}\right.\\
&\qquad\qquad
\left.\begin{matrix}
{}- \sin{2\omega}\sinh{2r}\sin(\theta + \psi + 2\phi_0) &
- \cos{2\omega}\sinh{2r}\sin(\theta + 2\phi_0)\\
{}- \cos{2\omega}\sinh{2r}\sin(\theta + 2\phi_0) &
\sin{2\omega}\sinh{2r}\sin(\theta - \psi + 2\phi_0)\\\cosh{2r} + \sin{2\omega}\sinh{2r}\cos(\theta + \psi + 2\phi_0) &
\cos{2\omega}\sinh{2r}\cos(\theta + 2\phi_0)\\
\cos{2\omega}\sinh{2r}\cos(\theta + 2\phi_0) &
\cosh{2r} - \sin{2\omega}\sinh{2r}\cos(\theta - \psi + 2\phi_0)
\end{matrix}\right).
\end{aligned}
\label{App:evolvedCov}
\end{equation}
Here $\Gamma_{\rm TMSS}$ denotes the covariance matrix of the input TMSS, as defined in Eq.~\eqref{eq:TMSSCOV}. Notably, the transformed covariance matrix $\Gamma_U$ remains independent of the parameter $\phi$ resulting that the corresponding quantum Fisher information $(\mathcal{H}^\Gamma)_{\phi\phi} = 0$ as shown in Eq.~\eqref{eq:QFIM_Gamma_TMSS}.

\subsubsection{Displacement derivative contribution of QFIM $\mathcal{H}^{\mathbf{d}}$}
The second term of QFIM $\mathcal{H}^{\bm{d}}$ in its full generality, i.e., for the arbitrary phases in the input state, is written as
\begin{align*}
\mathcal{H}^{\bm{d}}_{11} &= 4(|\alpha_1|^2 + |\alpha_2|^2)\cosh{2r} - 8|\alpha_1||\alpha_2|\sinh{2r}\cos(\theta - \beta_1 - \beta_2),\\
\mathcal{H}^{\bm{d}}_{12} &= 2(|\alpha_1|^2 - |\alpha_2|^2)\cosh{2r}, \\
\mathcal{H}^{\bm{d}}_{13} &= 2(|\alpha_1|^2 - |\alpha_2|^2)\cosh{2r}\cos{2\omega} - 2\sin{2\omega}\Big[ -2|\alpha_1||\alpha_2|\cos(\phi + \beta_1 - \beta_2)\cosh{2r} + (|\alpha_1|^2 \cos(\theta - \phi - 2\beta_1)\\
&\quad + |\alpha_2|^2 \cos(\theta + \phi - 2\beta_2))\sinh{2r} \Big], \\
\mathcal{H}^{\bm{d}}_{14} &= -8 |\alpha_1||\alpha_2| \cosh{2r} \sin(\phi + \beta_1 - \beta_2) + 4 \Big[ -|\alpha_1|^2 \sin(\theta - \phi - 2\beta_1) + |\alpha_2|^2 \sin(\theta + \phi - 2\beta_2) \Big] \sinh{2r}, \\
\mathcal{H}^{\bm{d}}_{22} &= (|\alpha_1|^2 + |\alpha_2|^2)\cosh{2r} + 2|\alpha_1||\alpha_2|\sinh{2r}\cos(\theta - \beta_1 - \beta_2), \\
\mathcal{H}^{\bm{d}}_{23} &= (|\alpha_1|^2 + |\alpha_2|^2)\cos{2\omega}\cosh{2r} + \Big[ 2|\alpha_1||\alpha_2|\cos{2\omega}\cos(\theta - \beta_1 - \beta_2) + (-|\alpha_1|^2 \cos(\theta - \phi - 2\beta_1)  \\
&\quad+ |\alpha_2|^2 \cos(\theta + \phi - 2\beta_2))\sin{2\omega} \Big]\sinh{2r}, \\
\mathcal{H}^{\bm{d}}_{24} &= -2(|\alpha_1|^2 \sin(\theta - \phi - 2\beta_1) + |\alpha_2|^2 \sin(\theta + \phi - 2\beta_2))\sinh{2r}, \\
\mathcal{H}^{\bm{d}}_{33} &= (|\alpha_1|^2 + |\alpha_2|^2)\cosh{2r}+ \sinh{2r}\big[ \sin{4\omega}\big(|\alpha_1|^2 \cos(\theta - 2\beta_1 - \phi) - |\alpha_2|^2 \cos(\theta - 2\beta_2 + \phi)\big) \\
&\quad + 2|\alpha_1||\alpha_2|\cos{4\omega}\cos(\theta - \beta_1 - \beta_2) \big],  \\
\mathcal{H}^{\bm{d}}_{34} &= -2 \cos{2\omega} (|\alpha_1|^2 \sin(\theta - \phi - 2\beta_1) + |\alpha_2|^2 \sin(\theta + \phi - 2\beta_2)) \sinh{2r}, \\
\mathcal{H}^{\bm{d}}_{44} &= 4(|\alpha_1|^2 + |\alpha_2|^2)\cosh{2r} - 8|\alpha_1||\alpha_2|\sinh{2r}\cos(\theta - \beta_1 - \beta_2).
\end{align*}
The QFIM $\mathcal{H}^{\bm{d}}$ allows at most two of its eigenvalues to scale as $O(N^2)$, so there will be only two parameters (functions of parameters) that achieve Heisenberg scaling. Here, we discuss a few cases by parametrizing the displacement amplitudes as $|\alpha_1|^2=\tau N_c, \; |\alpha_2|^2=(1-\tau) N_c$, where $\tau\in[0,1]$ is the weight of the displacement between the two input modes.
\paragraph*{\textbf{Case--1}} 
Consider the situation where the displacements at both input ports are equal, i.e., $\tau=1/2$, and the phase matching condition $\theta-\beta_1-\beta_2=0$. Under these constraints, $\mathcal{H}^{\bm{d}}$ yields only one eigenvalue that scales with $O(N^2)$. Also, the element $\mathcal{H}^{\bm{d}}_{22}$ is maximized.
\paragraph*{\textbf{Case--2}}
For an equal displacement amplitude ($\tau=1/2$), but relaxing the phase condition $\theta-\beta_1-\beta_2\neq 0$ significantly changes the structure of the QFIM $\mathcal{H}^{\bm{d}}$ and gives two eigenvalues that scale with $O(N^2)$. So we can estimate two parameters (functions of parameters, say $\Phi_1(\phi_0,\phi,\psi,\omega)$ and $\Phi_2(\phi_0,\phi,\psi,\omega)$ with Heisenberg-scaling sensitivity.
It is important to note that for the phase condition $\theta-\beta_1-\beta_2=\pi$, the parameter $\phi$ cannot be estimated with the Heisenberg scaling. 

\paragraph*{\textbf{Case--3}}
For unequal displacement amplitudes in the two modes, i.e., for any value of $\tau\neq1/2$, the QFIM $\mathcal{H}^{\bm{d}}$ always allows the two eigenvalues, scaling at $O(N^2)$, i.e., two parameters can be estimated simultaneously with Heisenberg scaling.
\begin{figure}
    \centering
    \includegraphics[width=0.5\linewidth]{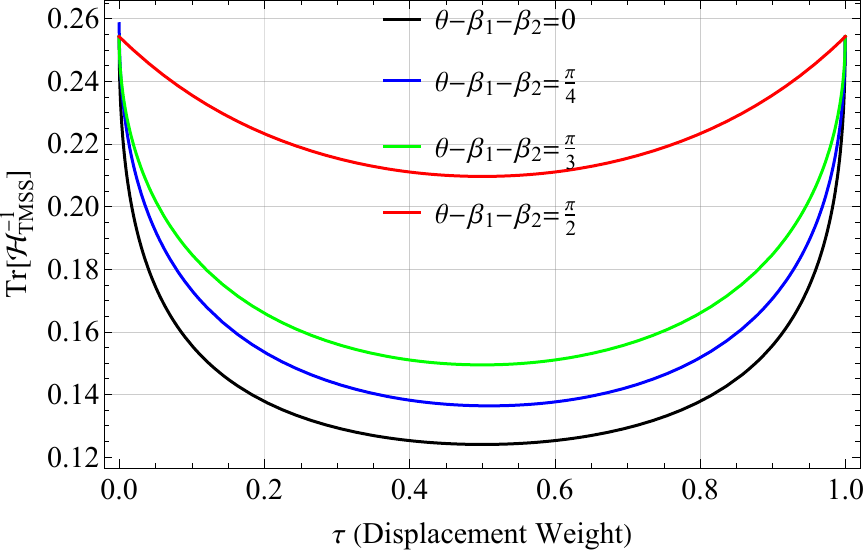}
    \caption{Plot of the scalar bound $\Tr[\mathcal{H}_{\rm TMSS}^{-1}]$ vs displacement weight $\tau$ by considering $N_c=N_s=N/2$. Here, $N=5$ and $\omega=\pi/4$.}
    \label{fig:displacement_weight}
\end{figure}
However, to achieve Heisenberg scaling in the estimation of all four parameters in total QFIM, $\mathcal{H}^{\bm{d}}_{22}$ must scale as $O(N^2)$ in the total average number of photons $N$. From the phase matching condition in $case-1$ and for balanced input displacements, we see that the element $\mathcal{H}^{\bm{d}}_{22}$ is maximized. Moreover, \figurename~\ref{fig:displacement_weight} shows that the optimal sensitivity is achieved by minimizing the scalar bound for optimal conditions $\theta-\beta_1-\beta_2=0$ and $\tau=1/2$.

Without loss of generality, one can choose the displacement and squeezed phases to be zero, i.e., $\beta_1=\beta_2=0$ and $\theta=0$. The QFIM $\mathcal{H}^{\bm{d}}$ for equal displacement amplitudes $\alpha_1=\alpha_2=\alpha$ then reads 
\begin{equation}
\mathcal{H}^{\bm{d}} = |\alpha|^2\left(
\begin{array}{cccc}
8 e^{-2r} & 0 & -8 e^{-2r} \sin{\omega} \cos{\omega} \cos{\phi} & 8 e^{-2r} \sin{\phi} \\
0 & 2 e^{2r} & 2 e^{2r} \cos{2\omega} & 0 \\
-8 e^{-2r} \sin{\omega} \cos{\omega} \cos{\phi} & 2e^{2r} \cos{2\omega} & 2(\cosh{2r}+\sinh{2r} \cos{4\omega}) & 0 \\
8 e^{-2r} \sin{\phi} & 0 & 0 & 8 e^{-2r} \\
\end{array}
\right).   
\end{equation}
\subsubsection{Total QFIM and QCRBs in terms of average number of photons}
We now express the total QFIM , given by $\mathcal{H} = \mathcal{H}^\Gamma + \mathcal{H}^{\bm{d}}$ in Eq.~\eqref{eq:QFIMmain} of the main text, in terms of the average number of squeezed photons $N_s=2\sinh^2{r}$ and the total number of displaced photons $N_c = |\alpha_1|^2 + |\alpha_2|^2=2|\alpha|^2$. Then by keeping only the leading contributions $O(N^2)$ for the Heisenberg scaling in $N_s$ and $N_c$, the total QFIM $\mathcal{H}$ given by
\begin{equation}
\mathcal{H}_{\rm TMSS} = 
\begin{pmatrix}
8 N_s^2 & 0 & 0 &0 \\
0 & 2N_c N_s & 2N_c N_s \cos{2\omega} & 0 \\
0 & 2N_c N_s \cos{2\omega} & 2 N_s^2 \sin^2{2\omega}+2N_cN_s \cos^2{2\omega} & 0 \\
0 & 0 & 0 & 8 N_s^2 \\
\end{pmatrix},
\end{equation}
by neglecting all orders smaller than $O(N^2)$, which do not contribute to the Heisenberg scaling in the precision, given by Eq.~\eqref{eq:QFTMSS}. 

Finally, the QCRB inequality defined in Eq.~\eqref{eq:QCRB}, implies that
\begin{equation}
\begin{split}
(\Delta\phi_0)^2&\geq(\mathcal{H}^{-1})_{11}=\frac{1}{8 N_s^2},\\
(\Delta\phi)^2&\geq(\mathcal{H}^{-1})_{22}=\frac{\frac{N_s}{N_c}+\cot^2{2\omega}}{2 N_s^2},\\
(\Delta\psi)^2&\geq(\mathcal{H}^{-1})_{33}=\frac{1}{2 N_s^2\sin^2{2\omega}},\\
(\Delta\omega)^2&\geq(\mathcal{H}^{-1})_{44}=\frac{1}{8 N_s^2}.\\
\end{split}
\end{equation}
These bounds indicate that Heisenberg scaling is achieved for all four parameters with parameter $\omega\in(0,\pi/2)$. The scalar bound plotted in \figurename~\ref{fig:figure2} and written by considering equal photons in squeezing and displacement, i.e., $N_s=N_c=N/2$, as 
\begin{equation}
    \Tr[\mathcal{H}_{\rm TMSS}^{-1}]\approx\frac{1}{N^2}\bigl(1+\frac{1}{\sin^2{\omega}\cos^2{\omega}}\bigr).
    \label{eqAp:scalarbound}
\end{equation}

\numberwithin{equation}{section}
\renewcommand{\theequation}{B\arabic{equation}}  
\section{Quantum Fisher Information Matrix for Input SMSS} \label{Appendix B}
\numberwithin{equation}{section}
\renewcommand{\theequation}{B\arabic{equation}}       
  \setcounter{equation}{0} 
In this Appendix, we will calculate the QFIM for the input of two SMSSs.~\eqref{eq:SMSS} and show that the QCRBs reach the Heisenberg scaling for all four parameters. The initial displacement and covariance matrix with squeezing phases $\theta_1=\pi/2$ and $\theta_2=0$ is given by
\begin{equation}
   \bm{d}= \sqrt{2}\begin{pmatrix}
\,|\alpha_1|\cos{(\beta_1)} \\
\,|\alpha_1|\sin{(\beta_1)} \\
\,|\alpha_2|\cos{(\beta_2)} \\
\,|\alpha_2|\sin{(\beta_2)}
\end{pmatrix},\;\quad 
    \Gamma=\frac{1}{2}\begin{pmatrix}
    \e^{2r_1}&0&0&0\\0&\e^{-2r_1}&0&0\\0&0&\e^{-2r_2}&0\\0&0&0&\e^{2r_2}
    \end{pmatrix}.
    \label{App:Gamma0SMSS}
\end{equation} 
We first calculate the covariance contribution of the QFIM $\mathcal{H}_{\rm SMSS}^\Gamma$ by evaluating $\Gamma_U=R\,\Gamma\,R^T$ from Eqs.~\eqref{App:R} and~\eqref{App:Gamma0SMSS} and yields the $4\times4$ matrix with matrix elements
\begin{equation}
\begin{split}
 ( \mathcal{H}_{\rm SMSS}^\Gamma)_{11}&=2(\cosh{4r_1}+\cosh{4r_2}-2),\; ( \mathcal{H}_{\rm SMSS}^\Gamma)_{12}= \cosh{4r_1}-\cosh{4r_2},\; ( \mathcal{H}_{\rm SMSS}^\Gamma)_{13}=\cos{2\omega}(\cosh{4r_1}-\cosh{4r_2}),\\
  ( \mathcal{H}_{\rm SMSS}^\Gamma)_{22}&= \tfrac12(\cosh{4r_1}+\cosh{4r_2}-2),\;( \mathcal{H}_{\rm SMSS}^\Gamma)_{23}=\tfrac12\cos{2\omega}(\cosh{4r_1}+\cosh{4r_2}-2),\\
   ( \mathcal{H}_{\rm SMSS}^\Gamma)_{33}&= \tfrac14\bigl[-4+(1+\cos{4\omega})(\cos{4r_1}+\cosh{4r_2})+4\sin^2{2\omega}(\cosh{2r_1}\cosh{2r_2}-\cos{2\phi}\sinh{2r_1}\sinh{2r_2}) \bigr] \\
    ( \mathcal{H}_{\rm SMSS}^\Gamma)_{44}&=2\bigl[\cosh{2(r_1-r_2)}+\cosh{2(r_1+r_2)}-2+2\sinh{2r_1}\,\sinh{2r_2}\,\cos(2\phi)\\
     ( \mathcal{H}_{\rm SMSS}^\Gamma)_{34}&=2\sin{2\omega}\,\sinh{2r_1}\,\sinh{2r_2}\sin(2\phi),\;\;(\mathcal{H}_{\rm SMSS}^\Gamma)_{14}=(\mathcal{H}_{\rm SMSS}^\Gamma)_{24}=0.
 \end{split}
\end{equation}

\begin{figure}
    \centering
    \includegraphics[width=0.5\linewidth]{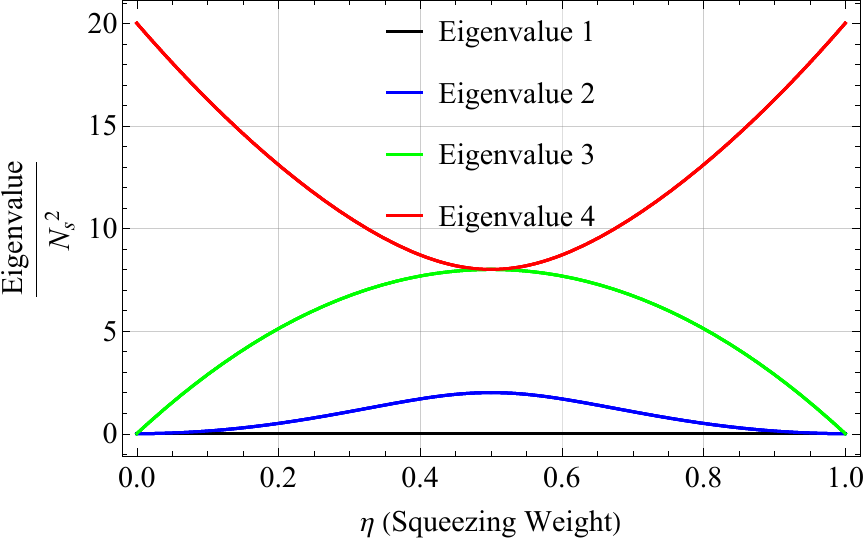}
    \caption{Eigenvalues of the QFIM $\mathcal{H}_{\rm SMSS}^\Gamma$ as a function of the squeezing weight $\eta\in[0,1]$. The squeezing weight
$\eta$ determines how the total mean number of squeezed photons $N_s$ is distributed between the two input modes, with $N_{s_1}=\eta N_s$ and $N_{s_2}=(1-\eta)N_s$. The plots illustrate that the first eigenvalue (in black) has a vanishing $O(N_s^2)$-coefficient for all \(\eta\); two eigenvalues are maximized at $\eta=1/2$, corresponding to equal squeezing in both modes. Here $N_s=3$, $\phi=0$, and $\omega=\pi/4$.}
    \label{fig:5}
\end{figure}
The above QFIM allows the Heisenberg-scaling estimation for at most three parameters. As shown in \figurename~\ref{fig:5}, one of the four eigenvalues has a vanishing $O(N^2)$-coefficient for all squeezing weights $\eta$ (i.e., it is zero at order \(O(N^2)\)). Whereas, two non-zero eigenvalues of $\mathcal{H}_{\rm SMSS}^\Gamma$ are maximized for equal squeezing in both inputs (i.e., $\eta=1/2$); however, the third non-zero eigenvalue is minimized. For simplicity, we choose both squeezings to be equal, $\eta=1/2$. The QFIM is then reduced to
\begin{equation}
\resizebox{1\textwidth}{!}{$
    \mathcal{H}^\Gamma_{\rm SMSS}=\begin{pmatrix}
 8 \sinh^{2 r} & 0 & 0 & 0 \\
 0 & \cosh{4 r}-1 & 2 \sinh^2{2 r} \cos{2\omega} & 0 \\
 0 & 2 \sinh^2{2 r} \cos{2\omega} & \frac{1}{2} \sinh^2{2 r} \left(\cos{4\omega}-2 \sin^2{2\omega} \cos{2 \phi}+3\right) & 2 \sinh^2{2 r} \sin(2\omega ) \sin{2\phi} \\
 0 & 0 & 2 \sinh ^2{2 r} \sin{2 \omega} \sin{2 \phi} & 8 \sinh^2{2 r} \cos^2{\phi} \\
    \end{pmatrix}.
    $}
    \label{Aeq:QFI_Gamma_SMSS}
\end{equation}
The matrix $ \mathcal{H}^\Gamma_{\rm SMSS}$ is singular and has rank three, indicating that only three independent parameters can be estimated when the input probe has zero displacement. This is directly analogous to the TMSS case, in which squeezing alone allowed the estimation of only three of four parameters. However, in contrast to the TMSS case, the QFIM $\mathcal{H}_{\rm SMSS}^\Gamma$ in Eq.~\eqref{Aeq:QFI_Gamma_SMSS} is not diagonal; its three nonzero eigenvalues still scale with $\sinh^2{2r}$, which leads to the Heisenberg scaling associated with three eigenvectors.

Further to calculate the Fisher term $\mathcal{H}^{\bm{d}}$, we can use the evolved displacement vector in Eq.~\eqref{App:evolved_d} as the displacement evolves similarly to the case of TMSS. However, in this case, to reach the Heisenberg scaling in all parameters, we choose the optimal values of the displacement phases to be $\beta_1=\beta_2=\pi/2$. 
For equal displacement amplitude $|\alpha_1|=|\alpha_2|=|\alpha|$, $\mathcal{H}^{\bm{d}}$ reads
\begin{equation}
  \resizebox{1\textwidth}{!}{$ \mathcal{H}^{\bm{d}}=|\alpha|^2\begin{pmatrix}
8 e^{-2r} & 0 & -4e^{-2r} \sin{2\omega}\sin{\phi} & -8e^{-2r} \cos{\phi} \\
0 &   2e^{-2r} &   2e^{-2r} \cos{2\omega} & 0 \\
-4 e^{-2r} \sin{2\omega} \sin{\phi} & 2 e^{-2r} \cos{2\omega} &   2\left(\cosh{2r}+ \sinh{2r} \left( \sin^2{2\omega} \cos{2\phi} -\cos^2{2\omega} \right) \right) & -4\sin{2\omega}\sin{2\phi}\sinh{2r} \\
-8e^{-2r} \cos{\phi} & 0 & -4\sin{2\omega}\sin{2\phi}\sinh{2r} & 8  \left( \cosh{2r}- \cos{2\phi}\sinh{2r} \right)
\end{pmatrix}.$}
\end{equation}
Now writing both terms of QFIM asymptotically in terms of average number of photons in squeezing $N_s=\sinh^2{r_1}+\sinh^2{r_2}=2\sinh^2{r}$ and in displacement $N_c=|\alpha_1|^2+|\alpha_2|^2=2|\alpha|^2$, as
\begin{equation}
  \mathcal{H}_{\rm SMSS}^\Gamma=  \begin{pmatrix}
8 N_s^2 & 0 & 0 & 0 \\
0 & 2 N_s^2 & 2 N_s^2 \cos{2\omega} & 0 \\
0 & 2 N_s^2 \cos{2\omega} & \frac{1}{2} N_s^2 \left( \cos{4\omega} + 2 \sin^2{2\omega} \cos{2\phi} + 3 \right) & 2 N_s^2 \sin{2\omega} \sin(2\phi) \\
0 & 0 & 2 N_s^2 \sin{2\omega} \sin{2\phi} & 8 N_s^2 \cos^2{\phi}
\end{pmatrix},
\end{equation}
\begin{equation}
   \mathcal{H}^{\bm{d}}= \begin{pmatrix}
\frac{2 N_c}{N_s} & 0 & -\frac{N_c \sin{2\omega} \sin{\phi}}{N_s} & -\frac{2 N_c \cos{\phi}}{N_s} \\
0 & \frac{N_c}{2 N_s} & \frac{N_c \cos{2\omega}}{2 N_s} & 0 \\
-\frac{N_c \sin{2\omega} \sin{\phi}}{N_s} & \frac{N_c \cos{2\omega}}{2 N_s} & 2 N_c N_s \sin^2{2\omega} \cos^2{\phi} &-2 N_c N_s \sin{2\omega} \sin{2\phi} \\
-\frac{2 N_c \cos{\phi}}{N_s} & 0 &-2 N_c N_s \sin{2\omega} \sin{2\phi} & 8 N_c N_s \cos^2{\phi}
\end{pmatrix}.
\end{equation}
The total QFIM $\mathcal{H}_{\rm SMSS}$ upto the order of $O(N^2)$ is written as
\begin{equation}
    \resizebox{1\textwidth}{!}{$ \mathcal{H}_{\rm SMSS}=\begin{pmatrix}
8 N_s^2 & 0 & 0 & 0 \\
0 & 2 N_s^2 & 2 N_s^2 \cos{2\omega} & 0 \\
0 & 2 N_s^2 \cos{2\omega} & \frac{1}{2} N_s \left[ 2 \sin^2{2\omega} \left( (N_c- N_s) \cos{2\phi} + N_c \right) + N_s (\cos{4\omega} + 3) \right] & 2 N_s (N_s - N_c) \sin{2\omega} \sin(2\phi) \\
0 & 0 & 2 N_s (N_s - N_c) \sin{2\omega} \sin(2\phi) & 8 N_s \left( N_c \sin^2{\phi} + N_s \cos^2{\phi} \right)
\end{pmatrix}.$}
\label{Aeq:totalQFIM_SMSS}
\end{equation}
Finally, the QCRB inequality using Eq.~\eqref{Aeq:totalQFIM_SMSS} is evaluated as
\begin{equation}
\begin{split}
(\Delta\phi_0)^2&\geq(\mathcal{H}^{-1})_{11}=\frac{1}{8 N_s^2},\\
(\Delta\phi)^2&\geq(\mathcal{H}^{-1})_{22}=\frac{3N_c+N_s-(N_c-N_s)(\cos{4\omega}+2\cos^2{2\omega}\cos{2\phi})}{8 N_c N_s^2\sin^2{2\omega}},\\
(\Delta\psi)^2&\geq(\mathcal{H}^{-1})_{33}=\frac{\left(N_s\cos ^2{\phi} +N_c \sin^2{\phi}\right)}{2N_c N_s^2 \sin^2{2\omega}},\\
(\Delta\omega)^2&\geq(\mathcal{H}^{-1})_{44}=\frac{(N_c-N_s) \cos (2 \phi )+N_c+N_s}{16 N_c N_s^2},\\
\end{split}
\end{equation}
which further simplifies for $N_c=N_s=N/2$ and reads
\begin{equation}
\begin{split}
(\Delta\phi_0)^2&\geq(\mathcal{H}^{-1})_{11}=\frac{1}{2 N^2},\\
(\Delta\phi)^2&\geq(\mathcal{H}^{-1})_{22}=\frac{2}{N^2\sin^2(2 \omega)},\\
(\Delta\psi)^2&\geq(\mathcal{H}^{-1})_{33}=\frac{2}{ N^2\sin^2(2 \omega)},\\
(\Delta\omega)^2&\geq(\mathcal{H}^{-1})_{44}=\frac{1}{2N^2}.\\
\end{split}
\end{equation}
These bounds indicate that Heisenberg scaling is achieved for all four parameters with parameter $\omega\in(0,\pi/2)$. This result is equivalent to the TMSS input, and gives a similar scalar bound in Eq.~\eqref{eqAp:scalarbound} written as
\begin{equation}
    \Tr[\mathcal{H}_{\rm SMSS}^{-1}]\approx\frac{1}{N^2}\bigl(1+\frac{1}{\sin^2{\omega}\cos^2{\omega}}\bigr).
\end{equation}

For analytical convenience, in the main text, we formulate this problem by choosing to perform a unitary transformation to couple the two modes of the input state. Specifically, the transformation we consider is $U_{\rm BS}=\e^{\i\frac{\pi}{4}(a_1^\dagger a_2-a_1a_2^\dagger)}$, which corresponds to interfering with the two input modes on a $50{:}50$ BS. Although this additional unitary rotation is not required to achieve the Heisenberg scaling for all three parameters, it simplifies the QFIM by diagonalizing it. Equivalently, one can see $U_{\rm BS}$ as an element of the general two-mode transformation $U$. It just reparameterizes the unitary evolution without affecting the ultimate precision limit. Moreover, one can also consider $U_{\rm BS}$ as part of probe preparation by applying it to two identical single‐mode squeezed vacuum states; one obtains a two‐mode squeezed vacuum state, which serves as the original input for our multiparameter estimation scheme, as shown in the main text.
\begin{equation}
    \Gamma_{\rm TMSS}\;=\;R(U_{\rm BS})\;\Gamma_{\rm SMSSs}\;R(U_{\rm BS})^T.
\end{equation}

\twocolumngrid 
\bibliography{mybib} 
\end{document}